\documentclass[preprint,12pt]{elsarticle}

\usepackage[a4paper,
            left=1.8cm,
            right=1.8cm,
            top=2.3cm,
            bottom=2.3cm]{geometry}

\usepackage[utf8]{inputenc}
\usepackage[T1]{fontenc}
\usepackage[british]{babel}

\usepackage{amsmath,amssymb}
\usepackage[version=4]{mhchem}
\usepackage{miller}

\usepackage{graphicx}
\usepackage{subcaption}
\usepackage{booktabs}
\usepackage{multirow}
\usepackage{adjustbox}

\usepackage[dvipsnames]{xcolor}
\usepackage{hyperref}
\hypersetup{
	colorlinks   = true,
	urlcolor     = darkorange,
	linkcolor    = RoyalBlue,
	citecolor    = RoyalBlue
}
\biboptions{sort&compress}
\usepackage{microtype}
\usepackage{siunitx}
\journal{Scripta Materialia}

\begin{document}

\begin{frontmatter}

\title{\raggedright Direct nanoscale observation of melting and solute redistribution in a hypoeutectic Al-Cu alloy with \textit{in situ} STEM}

\author[leoben1]{Martin Hasenburger\corref{cor1}}
\ead{martin.hasenburger@unileoben.ac.at}
\author[leoben2]{Rostislav Daniel}
\author[leoben1]{Phillip Dumitraschkewitz}
\author[leoben1]{Thomas M. Kremmer}
\author[leoben1]{Matheus A. Tunes}
\author[leoben1]{Stefan Pogatscher}

\cortext[cor1]{Corresponding author}

\address[leoben1]{Chair of Nonferrous Metallurgy, Department Metallurgy, Montanuniversit\"at Leoben, Franz-Josef-Stra\ss e 18, 8700 Leoben, Austria}
\address[leoben2]{Chair of Functional Materials and Materials Systems, Department Material Sciences, Montanuniversit\"at Leoben, Franz-Josef-Stra\ss e 18, 8700 Leoben, Austria}

\begin{abstract}
Melting and solidification of eutectic systems is a classical topic in physical metallurgy, yet the mechanisms at nanoscale are less investigated, due to experimental limitations in spatiotemporal resolution. The advent of \textit{in situ} STEM heating with MEMS technology has recently enabled investigation of eutectic behavior as a function of temperature, time and electrical resistivity. Using this methodology, we investigate the evolution of a nanocrystalline hypoeutectic Al--Cu alloy. Melting initiated in the hotter central region and propagated outward, with grain boundaries acting as preferred sites for eutectic liquid formation via Cu enrichment. The Al$_2$Cu phase melted prior to complete matrix melting. Liquid-state Cu redistribution over a distance of  \SI{258}{\micro\metre} -- several orders of magnitude beyond solid-state diffusion limits -- resulted in Al-rich rim accumulations and Cu enrichment at the outermost edge of the observed chip region. These observations are discussed in the context of classical predictions for melting of eutectic systems.
\end{abstract}


\end{frontmatter}


\noindent Al-Cu alloys are among the most studied age-hardenable aluminum systems, with microstructural evolution during thermal treatment, governing both processing performance and final properties~\cite{Chen2025}. When exposed to rapid heating, steep thermal gradients, or localized Joule heating, local melting and the associated redistribution of solute become decisive \cite{Dumitraschkewitz2022}. Such conditions inevitably arise in the context of technologically important processes, including additive manufacturing~\cite{Meier2017}, sintering~\cite{German2009}, strain-induced melt activation~\cite{BOLOURI2010}, casting~\cite{Stefanescu2015}, welding~\cite{Davies1975} and where transient partial melting along grain boundaries and within eutectic regions controls the resulting microstructure \cite{McKeown2016,Atkinson2005}. The melting behavior of nanocrystalline (NC) and ultrafine-grained (UFG) metallic materials is expected to differ from that of coarse-grained counterparts, yet remains experimentally largely uncharted. Molecular dynamics simulations and experiments have shown that the high grain-boundary volume fraction in nanometals causes boundary melting and a spatially distributed melting onset, with either depressed melting temperatures or lower onset sublimation, effects attributed to small grain sizes \cite{Noori2015,Coradini2024}. In multi-component NC/UFG systems, the shortened diffusion distances and elevated number of grain-boundary area per unit volume should dramatically accelerate eutectic liquid formation and solute partitioning relative to coarse-grained material. Heating with \textit{in situ} Scanning Transmission Electron Microscopy (S/TEM) studies of NC and UFG metals have so far addressed grain growth, precipitation, and solid-state phase transformations \cite{Legros2008, Coradini2023, Casu2024, Coradini2024}, leaving the melting regime and its nanoscale solute-redistribution consequences essentially unobserved directly. In hypoeutectic systems, grain boundaries are expected to act as preferred sites for eutectic liquid formation, locally enriching the lower-melting-point component before the matrix completely melts~\cite{Fischer2011,Straumal2024}. Once a liquid phase forms, additional transport mechanisms arise that can move solute over distances far exceeding solid-state diffusion \cite{scriven1960,Zhang2009}.
Electrothermal MEMS chips for \textit{in situ} TEM, onto which films can be deposited directly in the NC or UFG condition, offer a unique route to access this regime experimentally \cite{Protochips2020, Coradini2023}. 

Herein, we deposited a nanocrystalline hypoeutectic Al–Cu (Al$_{92}$Cu$_{8}$ at.\%) thin film directly onto a MEMS chip for a combined study with \textit{in situ} S/TEM, Energy Dispersive X-ray (EDX) spectroscopy, Selected Area Electron Diffraction (SAED), automated change-detection analysis, and simultaneous MEMS-based electrical biasing measurements to directly resolve the onset and spatial propagation of both the alloy and the Al$_2$Cu ($\theta$-phase) melting as well as the lateral redistribution of Al and Cu. The results demonstrate -- for the first time, to our knowledge, at nanoscale and millisecond spatiotemporal resolution -- that \textit{in situ} electrothermal S/TEM can directly probe the complex melting and solute-partitioning evolution of a hypoeutectic system at the nanoscale, providing relevant insights to a broader class of melting processes in engineering alloys.

\begin{figure}[t]
    \centering
    \includegraphics[width=0.95\linewidth]{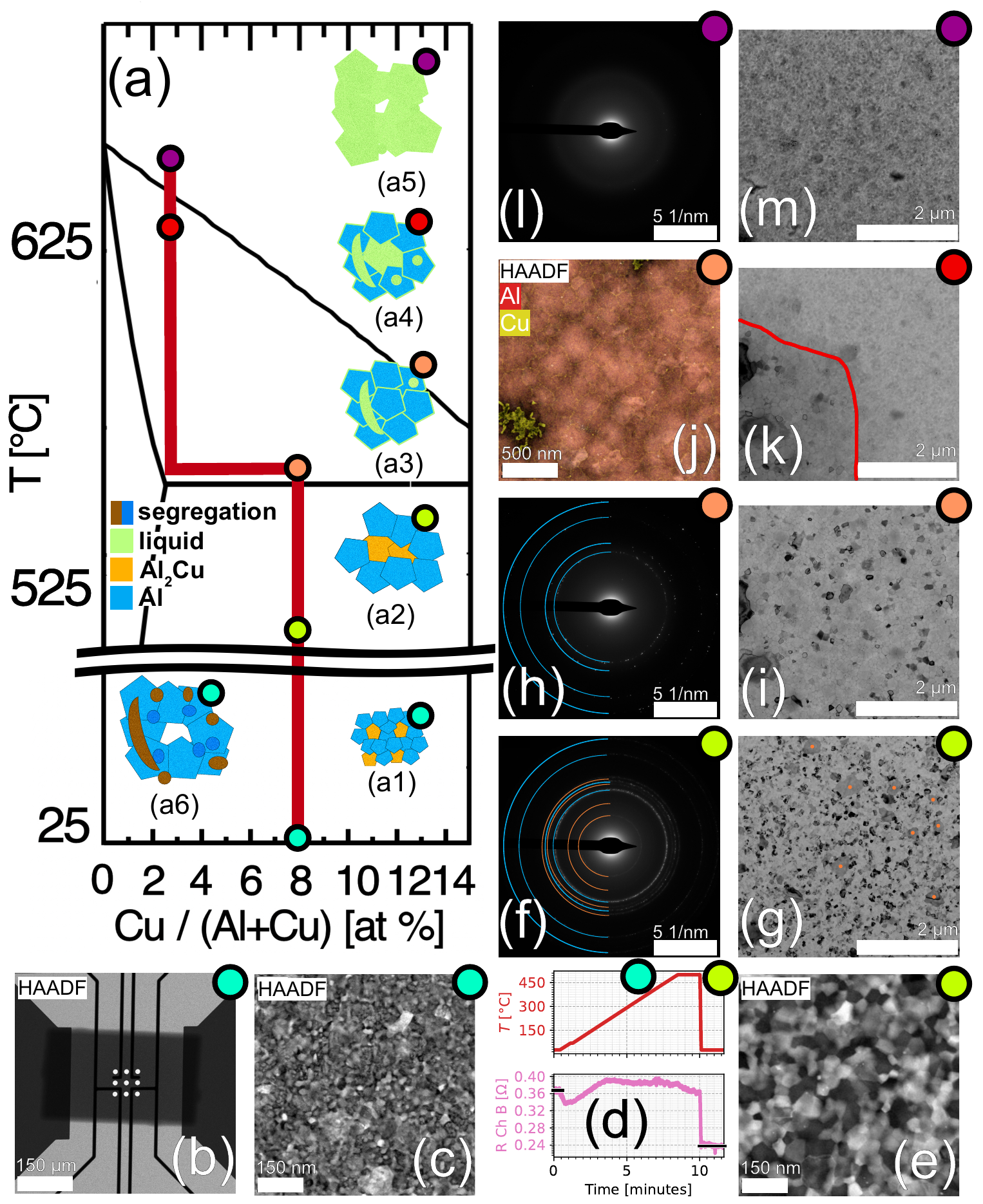}
    \caption{Overview of the phase transformation and transition sequence via heating. Cyan (cold) to violet (hot) dots. (a) Al-Cu phase diagram  and schematic thin-film state. After sputtering (a1), after ageing (a2), after eutectic segregation/melting (a3), after further melting of the residual thin film (a4), after film retraction/melting (a5), and after quenching (a6). (b) Overview LM-STEM and (c) detail STEM image of the as-deposited thin film on the chip membrane. (d) Electrical resistance of the thin film (channel B) during heating to \SI{500}{\celsius} and quenching. (e) Detail STEM image after heating. (f) SAED pattern at \SI{25}{\celsius}. (g) BF-TEM image of the Al-Cu thin film after five hours of ageing with $\theta$-phase marked by orange dots. (h) SAED pattern after heating above eutectic temperature, indicating melting of the $\theta$-phase. (i) BF-TEM image after melting of the eutectic. (j) EDX map above eutectic temperature, showing Cu segregation. (k) BFTEM image of the partially melted film (red border between melt and solid) (l) SAED pattern after melting. (m) BFTEM image after complete melting.}
    \label{fig:figure_1}
\end{figure}

Nanocrystalline hypoeutectic Al-Cu thin films with a thickness of \SI{100}{\nano\metre} were deposited by non-reactive magnetron sputtering in Ar at 0.2~Pa and room temperature. Pure 3~inch targets were operated at \SI{400}{\watt}, while an RF substrate bias of \SI{15}{\watt} was applied. The substrate holder rotated at \SI{1}{\hertz} to ensure a uniform film thickness. Sputter-deposition was carried out in a computer-controlled ultra-high-vacuum system (ATC 1800, AJA International) and further details of the deposition route are reported elsewhere \cite{Daniel2022Acta}. The thin film was deposited directly onto an electrothermal TEM chip mounted in a Protochips Fusion AX holder \cite{Protochips2020}. An overview STEM image of the sputtered chip is shown in Fig.~\ref{fig:figure_1}b. Microstructural and chemical characterizations were performed using a Thermo Fisher Scientific Talos F200X G2 operated at \SI{200}{\kilo\volt} in STEM-HAADF mode and with EDX at a microscope pressure of approximately $8 \times 10^{-6}$~Pa. Temperature control was performed using the Protochips Fusion Clarity software in combination with a Keithley 2450 source meter and a control time step of \SI{100}{\milli\second}. The nominal thin-film composition was Al$_{92}$Cu$_8$ (at.\%), as verified within the TEM using a Super-X EDX system. The chips have nine electron-transparent windows covered with only SiN for thermal transport. This SiN also contains oxygen. In addition to the SiN membrane, the chips contain a SiC heating element located outside the observation areas. The electrothermal chips are equipped with gold electrodes to obtain electrical resistance of the sample by four-point measurement.

An Al$_{92}$Cu$_8$ thin film was heated on the electrothermal chip until melting was noticeable in bright-field TEM (BFTEM) together with a pronounced change in the electrical signal. 
The red line in the phase diagram (Fig.~\ref{fig:figure_1}a) indicates the compositional changes in the main region of interest around the observation windows on the chip (Fig.~\ref{fig:figure_1}b) and shifts towards reduced Cu content, after passing \textit{solidus}, reflecting the outward diffusion of copper from the central region, resulting in a depletion of copper within the aluminum matrix. The microstructural evolution shown in Fig.~\ref{fig:figure_1} indicates that grain coarsening precedes the onset of melting -- see micrographs in Figs.~\ref{fig:figure_1}c, \ref{fig:figure_1}e, \ref{fig:figure_1}g, and \ref{fig:figure_1}i . However, the film remains within the UFG-regime up to temperatures close to melting. The process is schematically illustrated in Fig.~\ref{fig:figure_1}a1-a6. Fine aluminum grains and Al$_2$Cu ($\theta$-phase) after sputtering (a1) evolve to both larger grains and $\theta$-phase after aging (a2). Above the eutectic temperature, a copper-rich liquid forms at grain boundaries. Accumulation is preferentially observed at the outer regions of the sample (a3). The film starts to melt at the center of the chip (a4) with further heating and retraction of the central region towards the extremities (a5). After rapid quenching, the material concentrates at the outer regions and depleted zones remain in the observation area of the chip (a6). The schematically depicted process will be elaborated in the following with measured data: Fine/coarse grains and $\theta$-phase before/after heating to 500 °C can be seen in Fig.~\ref{fig:figure_1}c/e. STEM-HAADF images visualize the growth of $\theta$-phase (brighter parts). Al$_2$Cu ($\theta$-phases) were detected already in the as-deposited state and grew during subsequent annealing. The $\theta$-phase shows comparable size as the aluminum grains. 

During the \textit{in situ} annealing experiments, the variation of the film resistance was measured. The resistance curve before, during, and after heating is shown in Fig.~\ref{fig:figure_1}d. The resistance before and after is marked in black and indicates a 35~\% decrease from 0.37 to 0.24~$\Omega$, which is caused by a subsequent reduction of the volume fraction of GBs in the alloy~\cite{Palmer1987,Moraga2015}. The mean grain size grows from approximately 20 nm (c), to 50 nm (e), 150 nm (g) and 300 nm (i) during the annealing from RT to 500~°C. This is in consensus with literature, where after extended heat treatment, a grain has a maximum size two to three times the thin film thickness \cite{Barmak1996,Palmer1987}. Between the states shown in Fig.~\ref{fig:figure_1}e and Fig.~\ref{fig:figure_1}g, an aging treatment at \SI{500}{\celsius} for 5~h was performed. After this long-time treatment, the grains and precipitates increased 200~\% in size (from 50 to 150~nm). Individual grains increased in size by up to a factor of 40, comparing  the as-deposited state (20~nm to 800~nm) to the prior melting state. Fig.~\ref{fig:figure_1}g shows $\theta$-phase marked by orange dots, which dissolved during melting and were not detected after it as shown in Fig.~\ref{fig:figure_1}i. In Fig~\ref{fig:figure_1}f, showing the microstructure of the alloy at RT before the heating experiments where Debye-Scherrer (D-S) rings corresponding to Al and Al$_2$Cu are highlighted by blue and orange, respectively. Upon crossing the eutectic line, the orange rings disappear in Fig.~\ref{fig:figure_1}h. The diffraction data indicate that reflections associated with the $\theta$-phase melting, above the eutectic temperature and before complete melting of the Al-rich matrix, is in accordance to the expected behavior from the Al-Cu phase diagram. The melting of $\theta$-phase is also noted in the BFTEM sequence (marked by orange circles). 

Interesting at the direct observation at nanoscale is that the grain boundaries serve as Cu-rich facilitator regions to the eutectic melting procedure. The tendency of Cu to segregate to grain boundaries has been demonstrated previously~\cite{Fischer2011, Straumal2024}. This is evident from the EDX map in~\ref{fig:figure_1}j, where copper concentrated in the left bottom of the frame and at the grain boundaries. The melting sequence according to the BFTEM micrographs acquired during the \textit{in situ} experiment is visible from Fig.~\ref{fig:figure_1}i (no aluminum melt) to Fig.~\ref{fig:figure_1}k (partial melting with the transformation front indicated by the red line) to complete melting in Fig.~\ref{fig:figure_1}m. The grey contrast observed in Fig.~\ref{fig:figure_1}i is consistent with the presence of a Cu-rich liquid phase on top of the crystalline Al matrix; however, this cannot be unambiguously confirmed based on BFTEM alone. The pronounced grain contrast is related to Cu segregation to grain boundaries. Upon exceeding the \textit{liquidus} temperature, the diffraction rings associated with the Cu-rich phases disappear, while only the diffuse halo of the amorphous SiN membrane remains visible, suggesting that the Cu-rich phases have completely lost their crystalline structure through melting.

\begin{figure}[t]
    \centering
    \includegraphics[width=0.9\linewidth]{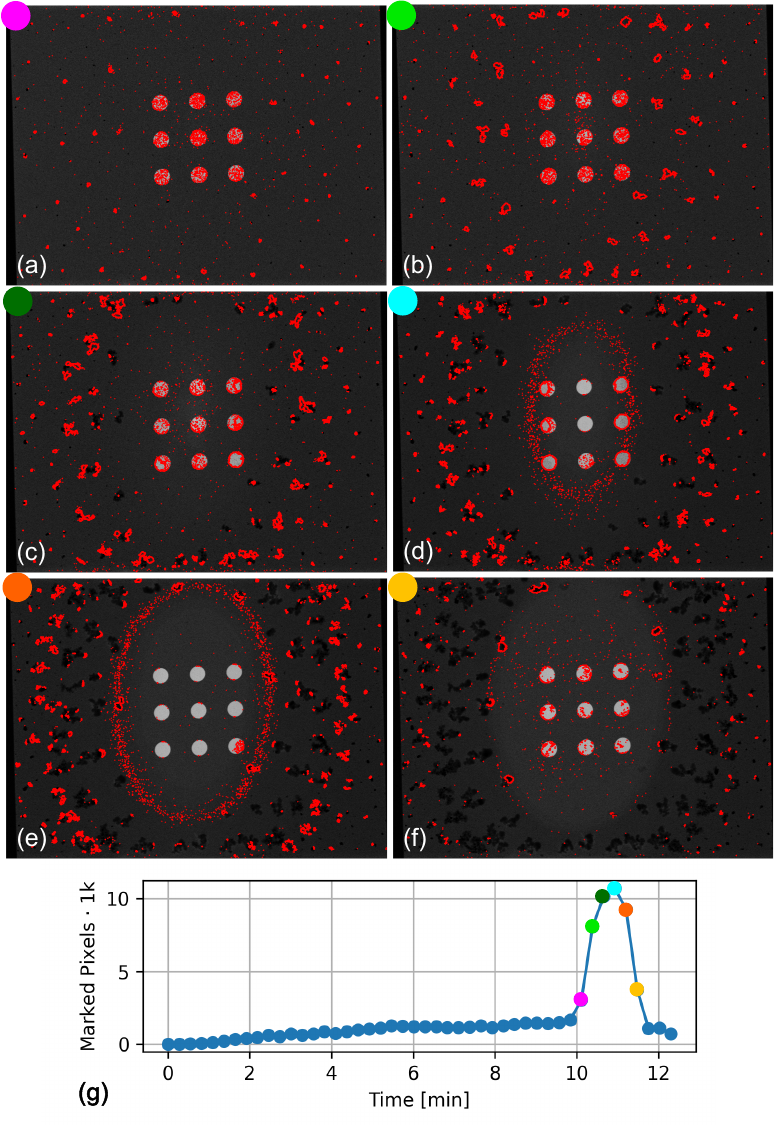}
    \caption{Low-magnification STEM overview of the melting process from Fig.~\ref{fig:figure_1}. The thin film retracts from the hotter central region towards the colder edge regions. Red overlays highlight changes between successive frames after noise averaging. The accompanying graph shows the temporal evolution of the detected pixel changes. Up to minute 10, the dominant processes are grain and precipitate growth; from minute 10 onwards, large-scale film contraction, melting and segregation become dominant. The frames a-f are marked by colored dots to specify the position in the time evolution (g).}
    \label{fig:figure_2}
\end{figure}

During heating, melting did not occur instantaneously across the full chip area. Instead, it started in the hotter central region, where typical samples are studied as the nine electron-transparent areas are intended as the observation area. Note that the melting propagated towards the outer regions of the chip, where the local temperature was lower, an effect easily overlooked when examining in high-magnification \textit{in situ} STEM studies. This spatially resolved progression of melting is depicted at low resolution in Fig.~\ref{fig:figure_2}, which captures solid and molten regions simultaneously. Regions of high pixel intensity differences are highlighted in red using an automated Python-based analysis. To visualize local changes in Fig.~\ref{fig:figure_2}, a simple frame-averaging and threshold-based change-detection routine was applied. The mean intensity of three consecutive frames was compared, and only grey-value changes between these means above 15 (0-255 scale) were marked. The images were additionally cropped to the thin-film region in order to exclude unrelated changes outside the area of interest. Figure~\ref{fig:figure_2} shows the start of the large-scale precipitation and grain growth on the edges, and the grain growth in the electron-transparent areas. This process continuously proceeds with time as shown in the following frame in Fig.~\ref{fig:figure_2}b. Fig.~\ref{fig:figure_2}c shows relatively large piles of aluminum and copper (black). The rapid diffusion at high temperatures resulted in pronounced flow of the material from the centre of the area of interest. This oval retraction is clearly highlighted by red contours in Fig.~\ref{fig:figure_2}d and \ref{fig:figure_2}e. The final frame in Fig.~\ref{fig:figure_2}f shows a large depleted area in the middle (with some residual material still left) and the thin film on the edges with large piles consisting of aluminium and copper. Figure~\ref{fig:figure_2}g shows the temporal evolution of the number of red pixels, which reflects the progression of melting. The frame~\ref{fig:figure_2}a is from minute ten, with equidistant in time (15~s differences) until the end of the video (frame~\ref{fig:figure_2}f). Note that the experiment shown in Fig.~\ref{fig:figure_2} was conducted using a chip without central electrical contacts to increase visibility of the changes, wherefore there are no contacts visible (black lines visible in Fig.~\ref{fig:figure_1}b).

\begin{figure}[t]
    \centering
    \includegraphics[width=\linewidth]{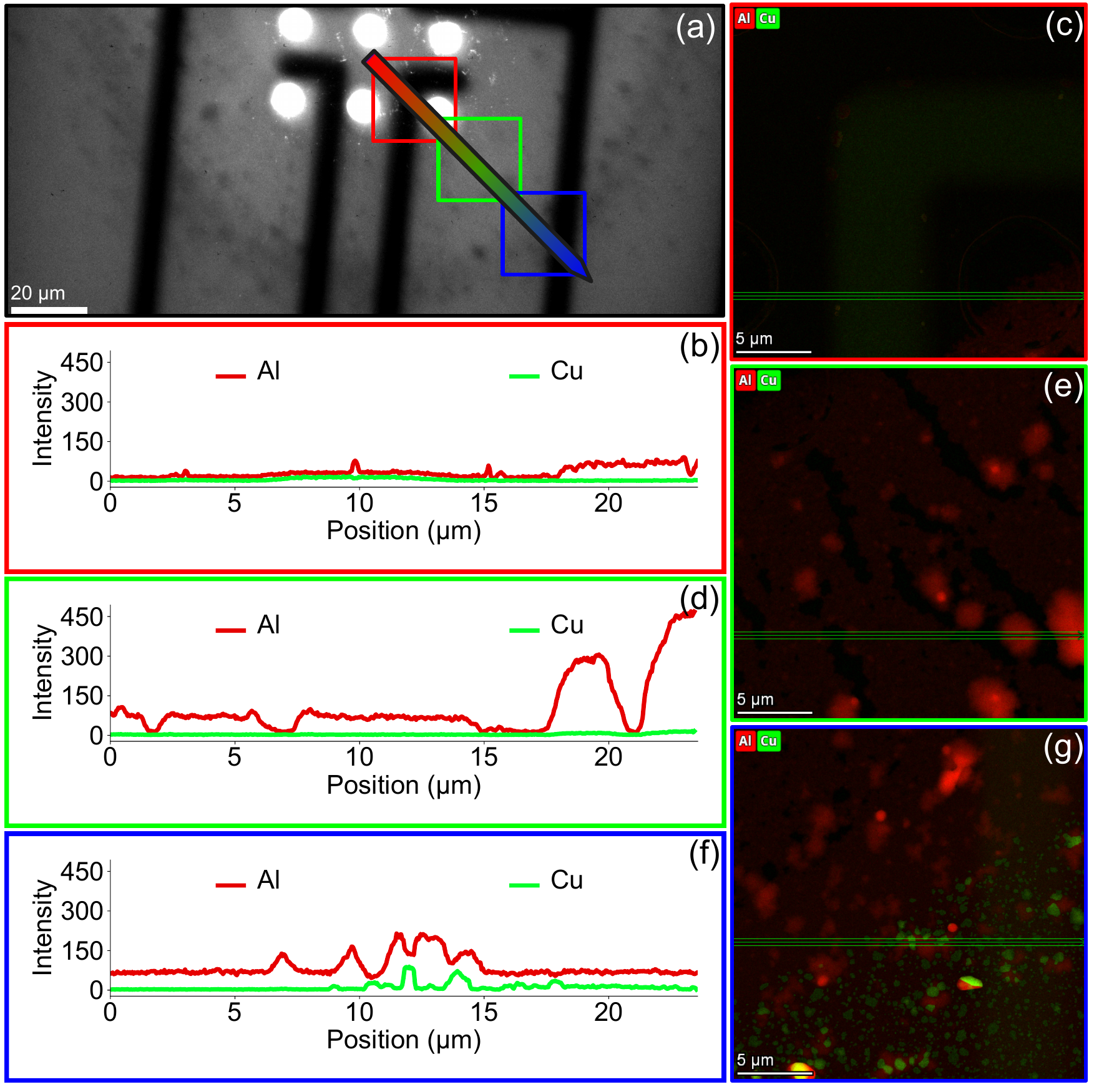}
    \caption{Chemical redistribution after melting. (a) Low-magnification STEM image of the final morphology. The arrow indicates the chip temperature gradient from hot (red) to cold (blue), as well as the flow direction of the melt. (b, c) Line profile and corresponding EDX map from a region in which most of the redistributed film accumulated in the lower right part of the frame, showing only minor Cu enrichment. (d, e) Line profile and corresponding EDX map from a region containing large Al-rich accumulations and channels depleted of material. (f, g) Line profile and corresponding EDX maps from a region with pronounced Al- and Cu accumulation.}
    \label{fig:figure_3}
\end{figure}

After the onset of melting, pronounced lateral redistribution of both Al and Cu was observed and properly quantified with STEM-EDX mapping. These results are shown in Fig~\ref{fig:figure_3}. The central region (red frame) became depleted in both Al and Cu, while material accumulated along the temperature gradient (going towards the blue frame), as can be seen in the low-magnification STEM micrograph in Fig~\ref{fig:figure_3}a. Figs.~\ref{fig:figure_3}b-g show STEM-EDX analyses from the three regions of interest highlighted by the red, green and blue squares in Fig.~\ref{fig:figure_3}a. The reported intensities correspond to background-corrected characteristic X-ray counts obtained from each EDX spectra. Each horizontal line profile was calculated by averaging ten adjacent vertical pixels to reduce statistical noise. As the eutectic melts, in the region of the red frame, both Al and Cu are observed depleted (Figs.~\ref{fig:figure_3}b-c). In the green region, only Al is observed with statistical significance (Figs.~\ref{fig:figure_3}d-e), whereas, in the blue region, both Al and Cu are observed, but the latter accumulates in the lower right corner (Figs.~\ref{fig:figure_3}f-g), suggesting that as a liquid phase forms, more extensive Cu transport takes place due to an earlier melting of the eutectic. 

\begin{figure}[t]
    \centering
    \includegraphics[width=1\linewidth]{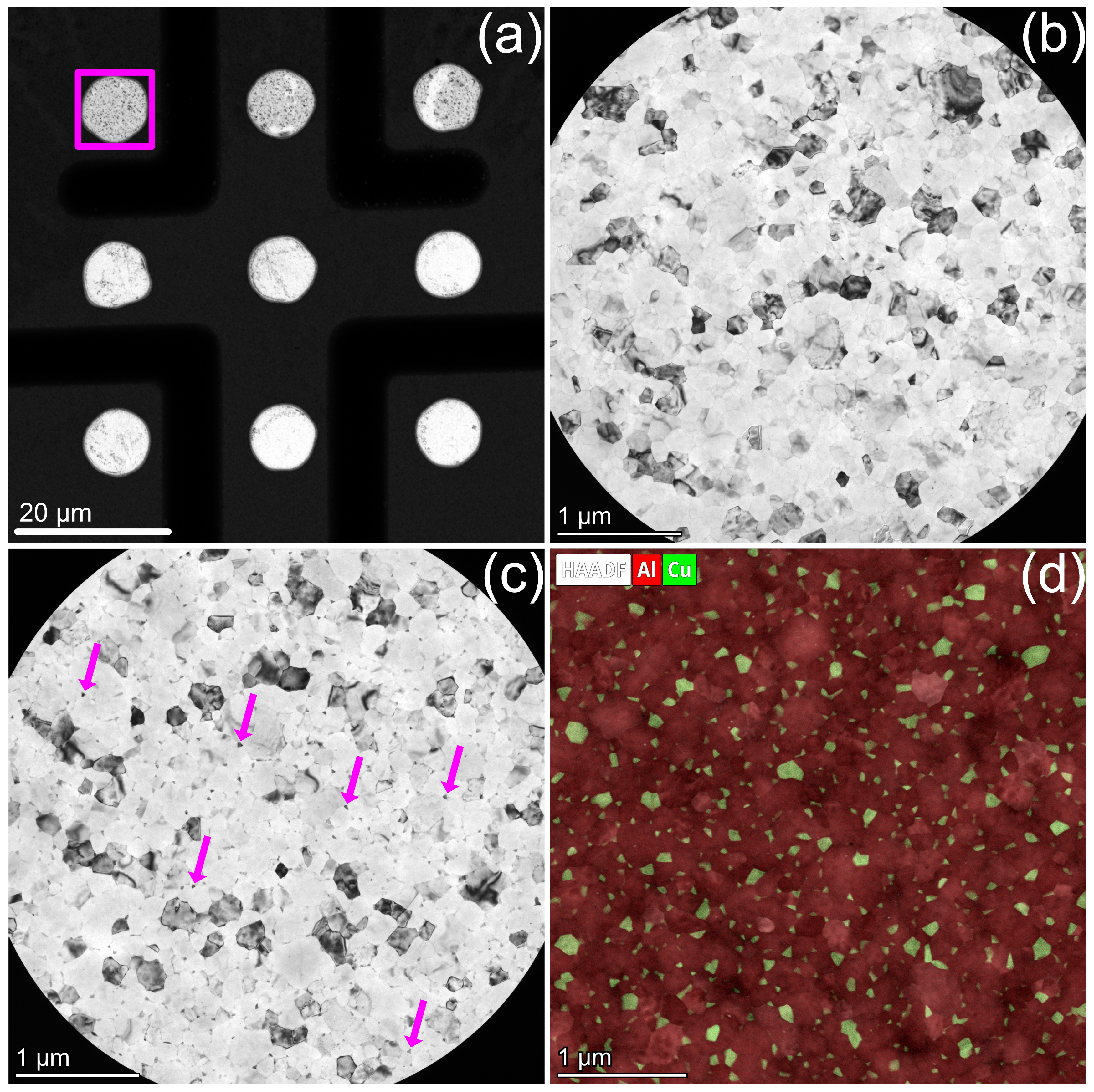}
    \caption{Post-ageing precipitates after quenching from the melting experiment. Precipitation occurs predominantly at grain boundaries, and some grains grow to dimensions comparable to the larger Al grains (approximately \SI{200}{\nano\metre}). (a) Overview STEM-HAADF. The precipitation area is marked by a pink square. (b) BFTEM with objective aperture, after quenching and before ageing. (c) BFTEM with objective aperture after halftime of ageing. (d) STEM-EDX after entire ageing time. The measured Cu level is approximately 2.16 at.\% considering the two-phase system of the entire area covered by the map in (d), determined by EDX, is consistent with re-precipitation from Cu previously dissolved in the Al-rich matrix.}
    \label{fig:figure_4}
\end{figure}

Additional evidence for the described process above is provided by the post-quench aging experiment shown in Fig.~\ref{fig:figure_4}. The same melting process as described in Fig.~\ref{fig:figure_2} was performed, but stopped at a similar frame as in Fig. \ref{fig:figure_2}d. The left upper electron-transparent window denoted by the magenta square in Fig.~\ref{fig:figure_4}a. is in a state, where the Cu segregated away (\textit{e.g.}, outside the field of view), leaving only dissolved Cu in the Al matrix represented by Fig.~\ref{fig:figure_4}b: in the phase diagram in Fig.~\ref{fig:figure_1}a, this denotes the middle region (a3) between \textit{solidus} and \textit{liquidus}. Fig.~\ref{fig:figure_4}b is a BFTEM micrograph after quenching taken from the magenta square region represented in Fig.~\ref{fig:figure_4}a: no $\theta$-phase precipitates are observed. Conversely, after half-time aging, Fig.~\ref{fig:figure_4}c shows intergranular precipitates. After full-time aging (\textit{i.e.}, 300~$^\circ$C for 30~min), EDX measurements on the quenched sample indicate a frame-global Cu concentration of approximately 2.16~at.\% as shown in Fig.~\ref{fig:figure_4}d. This value is below the maximum equilibrium solubility of Cu in aluminum (about 2.5~at.\% at 547 $^{\circ}$C) and significantly lower than the initial film composition of 8~at.\%. These observations are consistent with partial solute redistribution during melting. The post-mortem precipitates shown in Fig.~\ref{fig:figure_4}d further indicate that Cu remained dissolved in the Al-rich matrix after quenching and subsequently re-precipitated during aging. The preferential localization of these precipitates at grain boundaries suggests that both segregation and boundary-assisted precipitation contributed to the final microstructure \cite{Liu2019, Zhao2018, Hass2001}. 

The observed elemental redistribution can be rationalized by the combined action of non-uniform heating, thin-film retraction and phase partitioning. When the film melts locally, a temperature gradient is established between the hotter central region and the colder outer region. Niekie \textit{et al.} \cite{Niekiel2017} demonstrated that the temperature profile in this type of MEMS chip follows this behavior. For most liquid metals, surface tension decreases with increasing temperature

\begin{equation}
\frac{\partial \gamma}{\partial T} < 0,
\end{equation}

\noindent where $\gamma$ is the surface tension and T is the temperature. Consequently, a temperature gradient generates a surface-tension gradient that drives liquid flow from hotter regions of lower surface tension toward colder regions of higher surface tension. This phenomenon is known as the Marangoni effect~\cite{scriven1960,levich1962,davis1987,karbalaei2016}. At the same time, molten thin films are susceptible to capillarity-driven retraction and breakup in order to reduce the interfacial free energy \cite{thompson2012}. The experimentally observed depletion of the center and the formation of rim accumulations are consistent with such behavior.

Zhang \textit{et al.} \cite{Zhang2009} determined the diffusion coefficient of Cu in an Al melt with a comparable composition (Al$_{81}$Cu$_{19}$). By extrapolating the reported Arrhenius relationship to the temperature range relevant for the present experiments, a diffusion coefficient of $D_{\mathrm{Cu}} = 2.78 \times 10^{-10}$ m$^{2}$/s can be estimated. According to Fick's laws, an average distance for a self-diffusion species can be approximated as \cite{Fick1855,Shewmon2016}: 
\begin{equation}
x = \sqrt{2Dt}
\end{equation}

In Fig.~\ref{fig:figure_2}, the onset of Cu redistribution occurs after ten minutes, corresponding to approximately two minutes of melt propagation. Substituting this time into the diffusion relation yields a characteristic transport distance of 258~$\mu$m. This value is on the same order of magnitude as the lateral film dimension of about 240~$\mu$m. Such a transport distance strongly suggests that diffusion occurred in the liquid state, well in line with the observed Cu enrichment at the former grain boundaries melted at a3 in Fig.~\ref{fig:figure_1}. Diffusion coefficients of Cu in solid Al are typically in the order of $10^{-14}$ to $10^{-13}$ m$^{2}$/s~\cite{Du2003}, which would correspond to a diffusion distance of only about 5~$\mu$m over the same time interval. The experimentally observed redistribution therefore cannot be explained by solid-state diffusion alone. 

The electrical response changed markedly during heating and melting. Before melting, the resistance of the thin film decreased after heating (up to 500 °C) by 35 \%, which is consistent with grain growth and a subsequent reduction in electron scattering at grain boundaries. During heating in the solid state, the signal exhibits fluctuations due to the competing effects of Joule heating and microstructural evolution. However, the resistance increased again as melting set in by several orders of magnitude, until contact was lost due large-scale redistribution. 

In this work, using a hypoeutectic Al–Cu thin film as a model system, the onset of eutectic melting, subsequent solute partitioning, and large-scale material redistribution were visualized in real-time with \textit{in situ} STEM heating. The observations reveal how local melting, grain-boundary segregation, liquid-phase transport, and capillarity-driven film retraction interact during melting at the nanoscale. In particular, the combination of spatially resolved chemical analysis and electrical measurements allows the transition from solid-state coarsening to liquid-mediated transport to be identified directly. Beyond the specific Al–Cu system investigated here, the methodology provides a versatile platform for studying fundamental phase-transformation phenomena in a wide range of metastable nanostructured materials. Examples include precipitation and dissolution processes, segregation phenomena, eutectic reactions, local melting events, solidification pathways, and diffusion-controlled microstructural evolution. Such processes govern the behavior of alloys during additive manufacturing, welding, sintering, casting, microelectronic processing, and strain-induced melt activation. The present work, therefore, highlights the broader potential of electrothermal \textit{in situ} STEM for establishing direct structure–chemistry–property relationships during dynamic phase transformations.

\section*{Acknowledgments}
\noindent Funded/co-funded by the European Union (ERC, HETEROCIRCAL, 101124514). Views and opinions expressed are, however, those of the author(s) only and do not necessarily reflect those of the European Union or the European Research Council. Neither the European Union nor the granting authority can be held responsible for them. The research reported here was also supported by the Austrian Research Promotion Agency (FFG) within the project 3DnanoAnalytics (FFG No.~858040). Additionally, this work received funding from the innovation programme under the Marie Sklodowska-Curie grant agreement No. 897407

\section*{Declaration of competing interest}
\noindent The authors declare that they have no known competing financial interests or personal relationships that could have appeared to influence the work reported in this paper.


\section*{Data availability}
\noindent The raw data generated in this study have been deposited in the Zenodo repository under the accession code \href{https://doi.org/10.5281/zenodo.19706918}{doi:10.5281/zenodo.19706918}.

\section*{Code availability}
\noindent The Python scripts for plotting the electrical resistance and tracking the changes in the video are also deposited in the Zenodo repository under the accession code \href{https://doi.org/10.5281/zenodo.19706918}{doi:10.5281/zenodo.19706918}.



\bibliographystyle{elsarticle-num}
\bibliography{references}

\clearpage

\end{document}